\documentclass[11pt]{article}
\usepackage[T1]{fontenc}
\usepackage{lmodern}
\usepackage[margin=1in]{geometry}
\usepackage{amsmath}
\usepackage{array}
\usepackage{booktabs}
\usepackage{graphicx}
\usepackage{xcolor}
\usepackage{placeins}
\usepackage{float}
\usepackage{lineno}
\usepackage{xr-hyper}
\usepackage[hidelinks]{hyperref}
\makeatletter
\newcommand{\suppref}[2]{%
  \expandafter\ifx\csname r@S-#1\endcsname\relax #2\else\ref{S-#1}\fi}
\makeatother

\title{Closing the gap: A two-site MLC matching study for Varian TrueBeam Linear Accelerators}
\author{%
Michael Douglass\textsuperscript{1,2,*},
Corey Bridger\textsuperscript{1,2},
Mitchell Herrick\textsuperscript{1,2},
Joshua Southwell\textsuperscript{1,2},\\
Andrew Kennedy\textsuperscript{1,2},
Michael Barnes\textsuperscript{3,4},
Joerg Lehmann\textsuperscript{3,4,5}%
}
\date{}

\begin{document}
\maketitle
\begin{center}
\small
\textsuperscript{1}Department of Radiation Oncology, Royal Adelaide Hospital, South Australia, Australia\\
\textsuperscript{2}Adelaide University, School of Physical Sciences, Adelaide, South Australia, Australia\\
\textsuperscript{3}Calvary Mater Newcastle, Newcastle, NSW, Australia\\
\textsuperscript{4}University of Newcastle, Newcastle, NSW, Australia\\
\textsuperscript{5}University of Sydney, Sydney, NSW, Australia\\
\textsuperscript{*}Corresponding author: \href{mailto:michael.douglass@adelaide.edu.au}{michael.douglass@adelaide.edu.au}
\end{center}

\subsection*{Highlights}
\begin{itemize}
\item Beam-matched linacs had different MLC gap calibration states at baseline.
\item Four MLC gap-calibration methods tracked changes in the same direction.
\item Absolute leaf-position tests detected changes missed by routine picket fence QA.
\item Mechanical and dosimetric MLC calibration was harmonised across four matched linacs.
\item ArcCHECK measurements favoured the common planning-system gap value.
\end{itemize}

\begin{abstract}
\textbf{Purpose:} To characterise and harmonise multileaf collimator (MLC) gap calibrations across four beam-matched TrueBeams and assess a common planning-system model.

\textbf{Methods:} Four Millennium 120 TrueBeams were studied: TS1 (installed 2016), TS5 (2021), TS6 and TS7 (2024). Gap calibration was assessed by RayStation sweeping-gap measurement, feeler gauge, the EPID-based Stakitt Fence test, and a leaf-resolved Machine Performance Check (MPC). TS1 and TS5 were adjusted in service mode, TS5 also for MLC centreline offset. Ten clinical VMAT plans, and six acquired earlier on an unadjusted reference machine, were measured with an ArcCHECK 1220 against doses computed at 0.053 and 0.065~cm x-offsets in RayStation.

\textbf{Results:} The 2024 machines had wider baseline gaps. All four methods registered the TS1 and TS5 changes in the same direction and within 0.060~mm of predictions. After adjustment TS1 and TS5 lay within the TS6/TS7 sweeping-gap and Stakitt ranges but read wider on the feeler gauge. A conventional picket-fence test on TS5 registered no change. Raising the RayStation x-offset parameter increased mean local 3\%/2~mm pass rates from 92.3\% to 94.8\% on TS1 and 93.2\% to 95.6\% on TS5; all 24 plan-criterion pairs improved on the unadjusted machine.

\textbf{Conclusions:} Mechanical, EPID, and sweeping gap MLC gap measurements responded consistently to the MLC adjustment, moving TS1 and TS5 towards the local reference range. Every global 3\%/2~mm ArcCHECK measurement exceeded the TG-218 95\% limit.
\end{abstract}

\textbf{Keywords:} multileaf collimator (MLC); gap calibration; beam matching; TrueBeam; RayStation; quality assurance

\section{Introduction}

Accurate modelling of multileaf collimators (MLC) is essential for dose calculation in intensity-modulated radiotherapy (IMRT), volumetric modulated arc therapy (VMAT), and stereotactic techniques. Rounded leaf ends, transmission, tongue-and-groove effects, and systematic leaf-position offsets separate the geometric aperture encoded in a plan from the radiological aperture delivered, and matters most for small or highly modulated apertures: the fractional dose error from a systematic gap error grows as the mean leaf gap narrows. This is the mechanistic basis for the dosimetric leaf gap (DLG) concept~\cite{LoSasso1998}. Coherent bank-level errors are therefore far more consequential than random ones of similar magnitude, and errors in opposing-leaf aperture size carry greater clinical impact than a systematic shift of the aperture as a whole~\cite{Barnes2025PSQA}. Published tolerances vary with site, technique, and modulation: 0.3~mm has been proposed for systematic leaf-position error in IMRT at a 2\% deviation in target equivalent uniform dose, and approximately 0.6~mm per bank for systematic opening/closing errors in head-and-neck VMAT~\cite{Rangel2009,Oliver2010}. A separate planning study across head and neck, prostate, and lung stereotactic body radiotherapy put the per-bank error sufficient to cause major target-criterion violations at 0.2--0.7~mm, with mean target dose varying by 3.2--5.9\% per millimetre of per-bank displacement~\cite{Norvill2016}.

Routine measurement-based patient-specific QA (PSQA) is a poor detector of such errors: discriminating systematic MLC aperture errors at 3\%/3~mm required approximately \(\pm3\)~mm for a diode array~\cite{Hu2022}, and a 2\%/1~mm criterion detected 0.5~mm opening/closing displacements of each bank (a 1.0~mm aperture change) that 2\%/2~mm did not detect in stereotactic body radiotherapy~\cite{Kim2014}. AAPM Task Group 218 recommends global 3\%/2~mm gamma with a 95\% tolerance limit for modulated plans~\cite{Miften2018}.
The SEAFARER audit tests this directly rather than by simulation, embedding intentional delivery errors into plans that centres assess blind using their own routine PSQA. In its first implementation, seventeen erroneous plans across seven of seventeen institutions passed local PSQA despite raising spinal-cord dose by more than 5\%~\cite{Lehmann2022}. The subsequent multi-centre head-and-neck study, whose error set included both MLC banks opened or closed by 0.5 and 1.0~mm, found 21 of 44 centres passed at least one ``should-fail'' plan, with overall sensitivity and specificity of 79\% and 82\% and sensitivity ranging from 0 to 100\% between submissions using the same device. They found that a centre's performance was governed by protocol and analysis choices rather than by the detector~\cite{May2026}. 

A second, largely independent problem is the accuracy of the treatment planning system (TPS) MLC model, particularly its gap-sensitive terms. The conventional Eclipse (Varian Medical Systems, Palo Alto, CA, USA) MLC model represents these by a single DLG, which folds the radiological effect of the rounded leaf end together with any residual mechanical gap-calibration offset, so the two cannot be separated once commissioned. RayStation (RaySearch Laboratories, Stockholm, Sweden) represents these effects through separate model parameters: a leaf-tip width and a leaf-tip transmission describe the rounded leaf end, while a distinct MLC leaf-position x-offset shifts the effective radiological leaf-end position relative to its geometric position~\cite{RayPhysicsManual}. The optimal value need not equal the one returned by the vendor's own measurement procedure~\cite{Xue2018}, and in RayStation a 1~mm change in this parameter, for an Agility MLC, produced dose differences of up to 10\% in target and 15\% in spinal-cord dosimeters that conventional diode-array gamma analysis did not reliably identify~\cite{Koger2020}. The Eclipse parameter behaves similarly~\cite{Sjolin2016}. Saez et al.\ introduced a sweeping-gap formalism using synchronous and asynchronous dynamic gaps measured with a Farmer-type chamber, a measurement-based route to this correction~\cite{Saez2020}. A subsequent multi-institutional comparison found measured behaviour consistent across centres for a given physical MLC model while calculated doses departed from measurement by more than 10\% for some MLC--TPS combinations, and described the leaf-position shift these methods fit as a radiation field offset which ``absorbs the calibration offset introduced by the MLC controller''~\cite{Saez2023}. A high gamma pass rate therefore establishes neither mechanical calibration nor model accuracy, and because the fitted x-offset can partly compensate for the machine's mechanical state, its value requires local measurement.

Beam matching allows a plan to be delivered on any nominally equivalent linac without machine-specific re-planning, although the work that set out the concept found the vendor acceptance criteria insufficient to guarantee an optimal match~\cite{Sjostrom2009}. Those assessments emphasise depth doses, profiles, and output factors, and agreement of these broad characteristics does not demonstrate equivalence of MLC calibration. Comparing two TrueBeams that met most of the manufacturer's fine beam-matching criteria but failed the dose-difference criterion at \(3\times3\,\mathrm{cm}^2\), and which the authors therefore describe as only nominally beam matched, Guan et al.\ found the Eclipse DLG differed far more than any other parameter, by 53 to 85\% in relative terms across the four photon beams, while MLC transmission differed by less than 10\%~\cite{Guan2024}. One route to such a difference is the calibration procedure itself. The Varian installation procedure sets the opposing-leaf gap mechanically with feeler-gauge blades, a tactile endpoint reached without imaging or dosimetric readback, so different engineers, or the same engineer on different machines, may settle on gaps separated by an amount comparable to the tolerances above and service mode retains the current total gap value but not the adjustment history (Varian Medical Systems, personal communication). The centreline offset, the position of the closed-leaf junction relative to the collimator rotation axis, is set by a separate service-mode parameter that redefines the 0~cm leaf position and shifts both banks in the same direction. Of the two, the gap is the more consequential for delivered dose~\cite{Barnes2025PSQA} and is the primary focus of this work.

Routine MLC positional QA has historically relied on picket-fence-style tests, in which leaf positions are evaluated relative to the mean picket location. Automated EPID-based implementations generally report relative leaf-position error~\cite{Christophides2016}, leaving the conventional test largely insensitive to a systematic change in absolute leaf position or opposing-leaf gap. Barnes et al.\ developed the Stakitt Fence test, an improved EPID-based picket fence that references measured leaf positions to the collimator rotation axis and reports absolute individual leaf positions and opposing-leaf gap errors~\cite{Barnes2025}. Vendor-integrated tools such as the Varian Machine Performance Check (MPC) are complementary, deriving MLC-geometry metrics from automated EPID acquisitions~\cite{VarianMPC27,VarianMPC30,VarianMPC40}. A five-month comparison of the TrueBeam MPC Collimator Device Check with Stakitt fence on four linacs found bank-mean differences between the methods of up to 0.32~mm but agreement in detected introduced changes to within 0.18~mm~\cite{Barnes2025CDC}, direct evidence that two tests can track changes similarly while retaining method- and bank-specific offsets. Feeler-gauge measurement, by contrast, is a direct mechanical assessment, independent of any imaging or dosimetry system. Together these considerations motivate more than one independent MLC metric, and quantitative rather than purely qualitative routine tests.

This investigation began while commissioning two TrueBeams installed in 2024 at the satellite centre of our radiation oncology service. Compared with the service's two older TrueBeams (2016 and 2021), the newer machines showed differences in measured beam data for MLC-defined fields smaller than \(3\times3\,\mathrm{cm}^2\) and required a different effective leaf-position parameter during RayStation modelling. The vendor attributed this to an additional opposing-leaf gap calibration step introduced into the installation workflow in approximately 2019 (Varian Medical Systems, personal communication). Since the installation year is confounded with machine age, service history, and site in this cohort, that explanation was treated as a working hypothesis. The aim was to characterise and locally harmonise the gap-calibration states of four beam-matched Millennium-MLC TrueBeams so that a single RayStation beam model could serve all four, by (1) exploring cross-machine associations among physical feeler-gauge, dosimetric sweeping-gap, EPID-based Stakitt Fence, and MPC gap measurements; (2) determining whether service-mode adjustment of the two older machines moved their gap-sensitive measurements towards the local range of the newer machines; and (3) performing a post-calibration model-sensitivity analysis and documenting final-state PSQA on the two adjusted machines.

\section{Methods}

\subsection{Study design and machine cohort}

The radiation oncology service operates across two campuses, referred to here as the primary site and the satellite centre, which share a physics group, treatment planning system, and quality-assurance program. The harmonisation was undertaken so that one beam model could serve both sites and all four linacs. The primary cohort comprised four Varian TrueBeam linacs with Millennium 120 MLCs: TS1 (installed 2016) and TS5 (2021) at the primary site, and TS6 and TS7 (both 2024) at the satellite centre. TS7 was commissioned first of the 2024 machines and served as the initial local reference, with TS6 and TS7 jointly defining the reference range. TS1 and TS5 were adjusted to match the MLC calibrations of TS6 and TS7. Two further 2016 TrueBeams with high-definition MLCs (HD-MLCs), TS2 and TS3, were measured for context on method behaviour across a different MLC design, but their absolute gap metrics did not define the harmonisation range. Local designations run TS1 to TS7 with no TS4, a decommissioned machine that was not part of this study. The investigation comprised baseline measurement, service-mode adjustment of TS1 and TS5, repeat MLC measurement, and a post-calibration model-sensitivity and final-state PSQA assessment.

\subsection{RayStation sweeping-gap measurements}
\label{sec:sweeping_methods}

The RayStation MLC parameters were determined using the synchronous and asynchronous sweeping-gap method of Saez et al.~\cite{Saez2020}, with dose calculated in the clinical RayStation 2024A installation using the collapsed-cone algorithm on a 2~mm dose grid, for the clinically commissioned flattened 6~MV and 10~MV beams only. The x-offset was the primary gap-sensitive parameter. This parameter summarises the difference between measured and calculated MLC behaviour and appears in RayPhysics as the MLC collimator-calibration \(x\) offset, and is applied per leaf end, so that a given change in it corresponds to twice that change in opposing-leaf aperture~\cite{Saez2020}. The RayPhysics manual describes a positive offset as enlarging the field by that offset~\cite{RayPhysicsManual}, which is the displacement of each leaf end rather than the change in opposing-leaf aperture. The factor of two follows from the sweeping-gap formalism~\cite{Saez2020} and is supported by the measured responses of Section~\ref{sec:applied_response}. Parameter definitions and external reference values were taken from the RayStation 2024A RayPhysics manual~\cite{RayPhysicsManual} and the complete fitted parameter set from the current work is given in Supplementary Section~S2.

Measurements were performed using an NE~2571 Farmer chamber (Phoenix Dosimetry Ltd, UK) at the machine isocentre in Gammex solid water (Gammex Inc., WI, USA) at 90~cm source-to-surface distance and 10~cm depth, with the jaws at \(10\times10\)~cm\(^2\). The delivery methodology of Saez et al.~\cite{Saez2020} was followed using the DICOM sweeping-gap plans supplied by those authors. Synchronous sweeping gaps of 2, 5, 10, 20, and 30~mm were delivered. The asynchronous series used a 20~mm gap with leaf offsets sampled at 1~mm intervals from 0 to 10~mm and at 2~mm intervals from 10 to 30~mm. Leaf motion extended from a gap-centre position of \(-60\) to +60~mm, giving the 120~mm travel distance of the published formalism. MLC transmission fields for both banks and a \(10\times10\)~cm\(^2\) jaw-defined reference field completed the acquisition set.

\subsection{Physical feeler-gauge measurements}
\label{sec:feeler_methods}

Physical gap checks followed the  Varian installation procedure. A symmetric 1.0~mm MLC-defined field was set at isocentre in service mode and centred on the beam central axis. Calibrated metallic feeler-gauge blades were then passed between the opposing MLC banks with increasing thicknesses. The thickest blade that passed without perceptible friction was recorded. The recorded value was the measured physical gap at the MLC leaf plane, without projection or subtraction of the nominal gap. The leaf plane is 49.02~cm above isocentre, so the nominal 1.0~mm gap at isocentre corresponds to approximately 0.51~mm there. Each measurement was made by an in-house radiation engineer and independently cross-checked by a medical physicist, the same two individuals in every case, so no between-machine difference reported here can be attributed to a change of operator. One measurement set was acquired per machine and calibration state, so repeatability could not be estimated, and the effective resolution is limited by the available blade thicknesses of 0.05 mm.

\subsection{EPID-based Stakitt Fence measurements}
\label{sec:stakitt_methods}

Absolute MLC positions were evaluated using the Stakitt test of Barnes et al.~\cite{Barnes2025}, which reports absolute bank-position and opposing-leaf gap errors in place of the relative deviations of a conventional picket fence test. Separate plans for the Millennium 120 and HD120 MLCs were supplied by the authors of the original study, and the appropriate plan used for each machine. Plans were delivered at gantry \(0^\circ\) to minimise gantry-dependent MLC and EPID sag. Images were acquired with an AS1200 EPID at 100~cm source-to-imager distance using the 6~MV beam at 600~MU/min in the integrated dosimetry mode, with standard dark-field, flood-field, and off-axis profile corrections, and analysed with an in-house Python implementation of the published method. That implementation locates each leaf edge at the 50\% dose point of the profile across the edge, matching the criterion of the original paper~\cite{Barnes2025}, whereas MPC uses the steepest-gradient position~\cite{VarianMPC27,VarianMPC30,VarianMPC40}. A small method-dependent difference in the reported edge position of a rounded leaf end should therefore be expected. The EPID radiation field offset of the original implementation, which reconciles the light-field leaf position encoded in the plan with the 50\% radiation field edge~\cite{Barnes2025}, was set to zero throughout. The objective of the current work was to compare gap-calibration states between machines rather than to establish an absolute light- or radiation-field-defined leaf position. Omitting the correction displaces the reported values by a fixed amount for a given MLC model and energy, so it cancels in the pre/post comparisons and in the cross-machine comparison among the four Millennium-MLC machines on which the harmonisation rests. It does, however, contribute to the method-specific offset between Stakitt and MPC discussed in Supplementary Section~S1.

The central-axis reference came from two jaw-defined \(10\times10\)~cm\(^2\) fields acquired with the same beam and geometry, one at collimator \(270^\circ\) and one at \(90^\circ\). The centre of each was located from its 50\% intensity edges and the two averaged. A \(180^\circ\) change of collimator angle reverses the sign of any displacement between the jaw-defined field centre and the collimator rotation axis. The average returns the centre of collimator rotation, which the published method takes as the beam central axis~\cite{Barnes2025}. Referring the fence leaf positions to that coordinate is what makes the reported positions and gap deviations absolute rather than relative to the delivered field.

The leaf set analysed is set by the field, not by the MLC: positions are measured only along trajectories for which a leaf row is detected in the comb patterns at both sides of the image, and the Y-jaw setting of the supplied plan limits the imaged field to approximately \(\pm150\)~mm. On the Millennium 120 machines this returns the 40 central 5~mm leaf pairs and the five 10~mm pairs on each side, so the Stakitt bank means average approximately 50 of the 60 leaf pairs, against all 60 for MPC (Section~\ref{sec:mpc_methods}).

\begin{figure}[H]
\centering
\begin{minipage}[t]{0.48\textwidth}
\centering
\includegraphics[width=\linewidth]{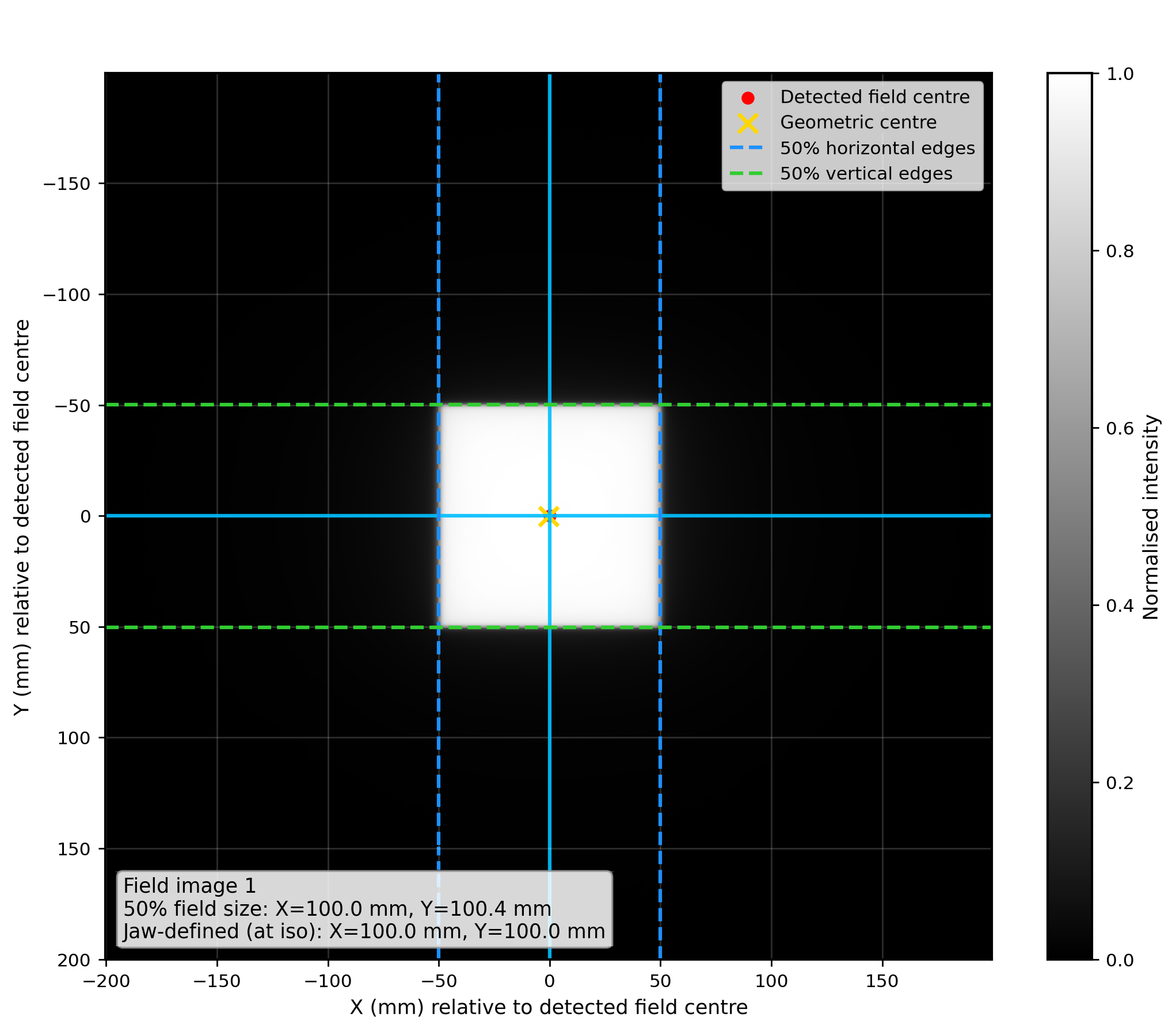}\\[-0.1em]
\small (a) Central-axis reference field
\end{minipage}\hfill
\begin{minipage}[t]{0.48\textwidth}
\centering
\includegraphics[width=\linewidth]{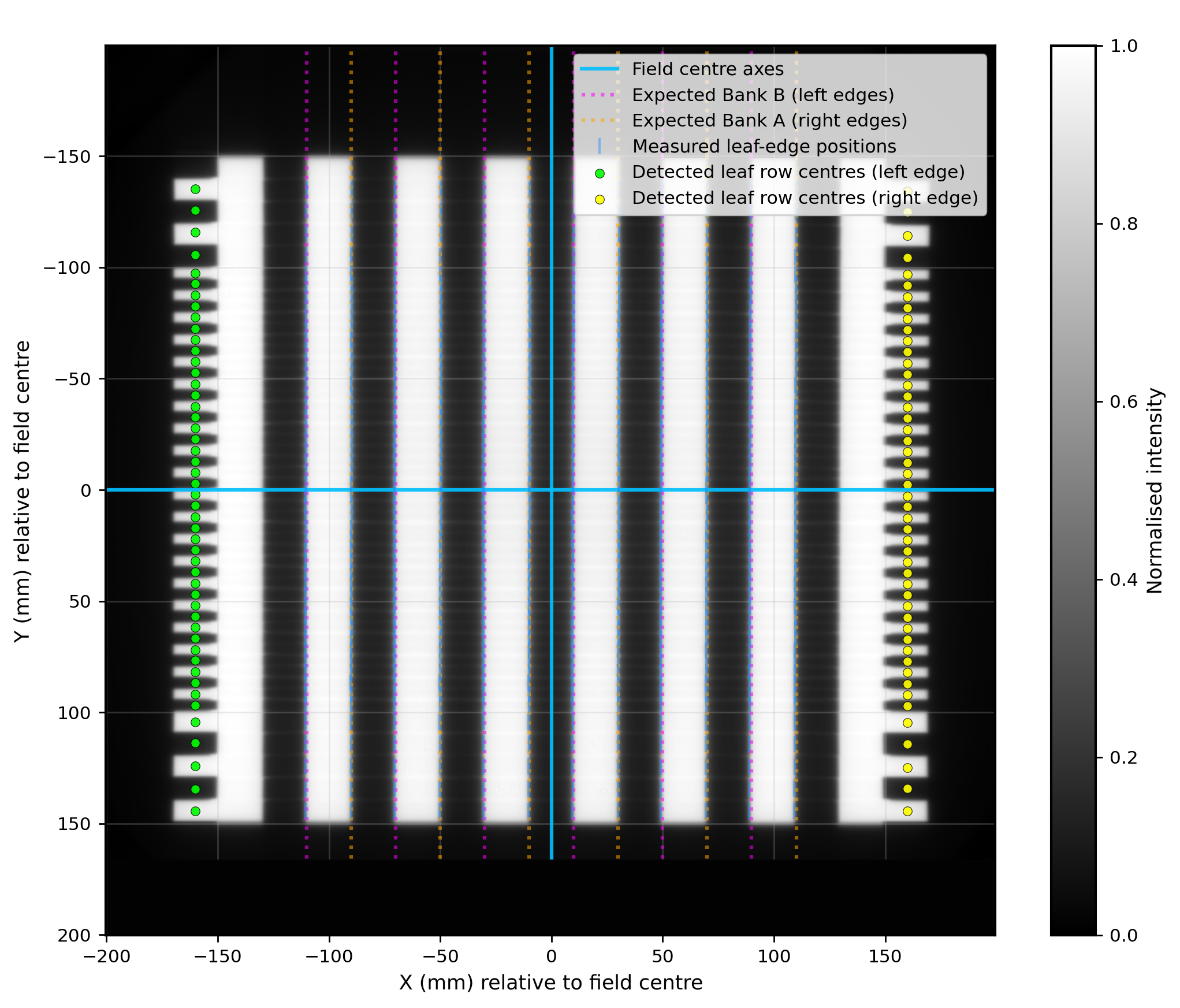}\\[-0.1em]
\small (b) Stakitt Fence field
\end{minipage}
\caption{Stakitt Fence analysis. (a) One of the two jaw-defined \(10\times10\)~cm\(^2\) reference fields, showing the detected field centre and 50\% intensity edges; the centres of the collimator \(270^\circ\) and \(90^\circ\) fields were averaged to give the centre of collimator rotation used as the reference for panel (b). (b) Delivered Stakitt fence field, showing the field-centre axes, the expected Bank~A and Bank~B leaf-edge positions (dotted lines), the measured leaf-edge positions, and the leaf-row centres detected from the comb patterns at the left and right of the field, which define the leaf trajectories along which the edges are measured.}
\label{fig:stakitt_field}
\end{figure}

Gap deviation was defined as \(\Delta g_{\mathrm{Stakitt}} = g_{\mathrm{measured}}-g_{\mathrm{planned}}\), so positive values indicate a radiological gap wider than planned. The Stakitt field comprised six strips, each of nominal 2~cm opposing-leaf gap and spaced 4~cm centre to centre, so that the closed region between adjacent strips was also 2~cm wide. The analysis is illustrated in Figure~\ref{fig:stakitt_field}.

\subsection{Machine Performance Check measurements}
\label{sec:mpc_methods}

Varian MPC was run at 6~MV on all machines. MPC versions follow the TrueBeam platform version and differed across the cohort: TS1 and TS5 ran TrueBeam~2.7, TS7 ran 3.0, and TS6 ran 4.0. The corresponding reference guides are the April 2024 revisions for versions 2.7 and 3.0 and the November 2025 revision for version 4.0. MPC determines MLC leaf-positioning accuracy from EPID images of an alternating-leaf pattern, measuring each leaf tip's distance from the MLC centreline, defined as the line through the centre of MLC rotation perpendicular to the leaf edges. The reported leaf-position values are adjusted by the configured MPC leaf-gap offset~\cite{VarianMPC27,VarianMPC30,VarianMPC40}. The reported offsets therefore depend on the MPC configuration and should not be interpreted directly as absolute physical leaf positions or opposing-leaf gaps. The acquisitions comprised all 60 leaf pairs on the Millennium machines and leaves 2 to 59 on the HD120 machines, the leaf set following the check that is run rather than the MLC model~\cite{VarianMPC27,VarianMPC30,VarianMPC40}. The HD120 acquisitions were standard beam and geometry MPC checks rather than enhanced MLC MPC checks and are contextual only.

Two opposing-leaf gap metrics were derived. The primary metric, \(g_{\mathrm{MPC}}\), was computed from the leaf-resolved arrays in the exported \texttt{Results.xml} file as the unweighted mean across both approach directions and, on the position-resolved software versions, across all MLC positions. A MLC centreline metric was taken as the half-difference of the two bank values. The MPC bank summaries in the exported \texttt{.csv} file were not used for this analysis as the per-leaf values underlying each is a signed, magnitude-selected result rather than a direction mean. A second metric, \(g_{\mathrm{CSV}}\), was formed from those summaries and is reported alongside throughout. This value is what a user obtains from the standard export and the quantity against which the external Stakitt comparison of Barnes et al.\ was made~\cite{Barnes2025CDC}, so the divergence between the two isolates the effect of the magnitude-selection rule (Section~\ref{sec:mpc_csv_vs_xml}). MPC is configured with the user-added dosimetric gap alone, not the total gap displayed in the MLC service window; that value (\texttt{MLCLeafGapOffset}) was read from each data file rather than assumed. The MPC leaf gap offset value was 0.025~cm on TS1, 0.030~cm on TS3, and zero elsewhere, and was left unchanged through the adjustment, which is why the MPC data retain sensitivity to it. Both metrics are defined in full, with their sign conventions and the reasoning behind them, in Supplementary Section~S1, together with the extraction procedure, the reconstruction of the exported summaries, the direction-assignment rule and its verification, and the differences between software versions.

\subsection{Gap-calibration and centreline-offset adjustment}

Following review of the baseline MLC measurements on TS1, TS5, TS6 and TS7, the gap calibration of TS1 and TS5 was increased in TrueBeam service mode. TS5 reached a final displayed total gap of 0.06~cm through a short titration and TS1 being set directly to 0.05~cm. Both linacs had previously displayed 0.00~cm. Adjustment stopped when repeat EPID measurements placed TS5 within approximately 0.1~mm of TS6 and TS7 and near their midpoint, the feeler gauge confirming the direction and approximate magnitude of each step rather than defining the endpoint. This stopping rule was applied during the session and was not prespecified. A systematic TS5 centreline-offset correction of +0.02~cm was also applied after the EPID centring field and MPC history indicated a long-standing misalignment of approximately 0.4~mm between the banks, equivalent to a 0.2~mm shift of their common centerline. No centerline adjustment was required on TS1. The gap was increased rather than decreased to avoid increasing opposing-leaf collision risk, consistent with vendor recommendations.

During the TS5 MLC adjustment process, the displayed total gap was first increased to 0.04~cm, matching the TS7 installation setting. This proved insufficient, because the TS5 baseline was narrower than the TS1 baseline. It was then increased to 0.07~cm, at which the feeler gauge read approximately 0.65~mm at the leaf plane against 0.48~mm on TS7, and therefore reduced to the final 0.06~cm. Varian support clarified (Varian Medical Systems, personal communication) that the service interface reports only the current total gap value, neither distinguishing the installer calibration from subsequent dosimetric adjustments nor retaining any history of them. The 0.5~mm physical separation at the leaf plane used during installation is accounted for inversely, so a planned Bank~A/Bank~B position of 0~cm still corresponds to complete centreline closure.

The MLC centerline check used a 6~MV beam and two opposed MLC defined half-fields at collimator \(270^\circ\) and \(90^\circ\), the EPID imager at 100~cm source-to-imager distance in Dosimetry mode, averaged in ImageJ to produce each panel of Supplementary Figure~\suppref{fig:ts5_centreline}{S1}. Junction-intensity asymmetry was assessed visually and checked against the Stakitt bank-offset results. This analysis lay outside the main scope of the current work and did not contribute to the gap measurement or calibration endpoint. After adjustment, the EPID centring field and a conventional picket-fence test served as qualitative sanity checks, and the sweeping-gap, feeler-gauge, Stakitt, and MPC measurements were repeated with the baseline acquisition settings. 

\subsection{Post-calibration model-sensitivity and clinical QA assessment}

Once the MLCs had been calibrated on TS1 and TS5, ten retrospectively selected clinical VMAT plans, six 6~MV and four 10~MV, spanning chest wall, head and neck, breast, lung, anal canal, rectum, and prostate treatments, were measured on both adjusted machines with an ArcCHECK Model 1220 cylindrical diode array (Sun Nuclear Corporation, Melbourne, FL, USA) and analysed in absolute dose. All ten were conventionally fractionated and none was stereotactic, so their aperture sizes and modulation are those of routine practice rather than the small-field regime in which gap sensitivity is greatest. Each measurement was compared with dose calculated in RayStation 2024A on a 2~mm dose grid at MLC leaf-position x-offsets of 0.053 and 0.065~cm, the former the value carried by the beam model then in clinical use on TS6 and TS7. The 0.065~cm setting was retained in the final common model (TB\_SD), with external plausibility comparisons against Saez et al.\ and the RayStation 2024A RayPhysics manual~\cite{Saez2023,RayPhysicsManual}.

The clinical TB\_SD model did not adopt the locally fitted parameter set in full. The transmission, leaf-tip width, tongue-and-groove width, and leaf-tip transmission were vendor reference values checked during the commissioning of TS6 and TS7, and only the x-offset was tuned locally (Supplementary Section~S2). Because the same ten plans were used both to compare the two x-offsets and to describe the retained model this is a model-sensitivity analysis rather than independent validation. Local- and global-gamma pass rates were calculated with the manufacturer's analysis software, version 6.6, at 3\%/3~mm, 3\%/2~mm, 3\%/1~mm, 2\%/2~mm, and 2\%/1~mm with a 10\% low-dose threshold and the \emph{Apply Measurement Uncertainty} option unchecked, so the pass rates are not inflated by a measurement-uncertainty allowance. Global results were recorded only for the retained model in this ten-plan dataset, so the paired comparison between the two x-offsets used the local criteria alone. AAPM TG-119 benchmark plans~\cite{Ezzell2009} were also delivered in the same sessions but the results are not discussed in the current study.

A second, separate dataset provided a cross-check. Six clinical VMAT plans, identified here as Plans~11--16, distinct from the ten above and likewise conventionally fractionated, had been measured on TS7 during that machine's 2024 commissioning validation, when the pre-harmonisation model at 0.053~cm was in clinical use. Those measurements were retained, the same plans recalculated with TB\_SD at 0.065~cm and re-analysed against the original measurements, so the calculation was the only quantity that changed. TS7 was never mechanically adjusted, and neither these plans nor this machine contributed to selecting the x-offset. Local and global pass rates were available for this dataset at 2\%/2~mm and 3\%/2~mm only.

\subsection{Statistical analysis}

Because the primary cohort contained only four Millennium-MLC machines and TS1/TS5 contributed paired pre/post observations, analyses were descriptive. Exploratory Pearson correlation coefficients were calculated for the feeler-gauge, 6~MV sweeping-gap, Stakitt, and MPC gap metrics, the last in both derived forms, over the six Millennium-MLC observations and over all eight including the two HD-MLC observations; the eight-observation values are those annotated on Figure~\ref{fig:mlc_method_comparison}. MPC values used in the cross-machine correlation were those with the configured offset restored, the unrestored values also being reported (Supplementary Section~S1). No hypothesis tests or confidence intervals were reported, the observations being neither independent nor numerous enough to support inferential statistics, and correlation was read only as an exploratory description of association, not as evidence of agreement or absence of bias. ArcCHECK values were analysed plan by plan, paired changes between the two x-offsets summarised by machine, energy, and across all 20 machine-plan pairs, and the TS7 cross-check the same way across its six plans and four criteria.

\section{Results}

\subsection{MLC gap measurements}
\label{sec:gap_measurements}

Table~\ref{tab:gap_measurements} gives the four comparison metrics for all machines and calibration states, the MPC metric in both derived forms, each as measured and after the configured gap offset has been restored.

\begin{table}[H]
\centering
\caption{MLC gap-calibration measurements (mm). The four metrics are not on a common scale and should not be read across a row as equivalents: feeler-gauge values are physical gaps at the MLC leaf plane, and the sweeping-gap column is the 6~MV fitted x-offset, a per-leaf-end quantity, so a given change in it corresponds to twice that change in opposing-leaf aperture (Section~\ref{sec:applied_response}). TS2 and TS3 are HD-MLC contextual observations.}
\label{tab:gap_measurements}
\small
\resizebox{\textwidth}{!}{%
\begin{tabular}{llrrrrrrr}
\toprule
& & & & & \multicolumn{2}{c}{MPC, XML-derived} & \multicolumn{2}{c}{MPC, CSV-derived} \\
\cmidrule(lr){6-7}\cmidrule(lr){8-9}
Machine & State & Feeler gauge & Sweeping gap (6X) & Stakitt Fence & Measured & Restored & Measured & Restored \\
\midrule
TS1 & Pre & 0.35 & 0.38 & -0.014 & -0.410 & -0.160 & -0.73 & -0.48 \\
TS1 & Post & 0.55 & 0.62 & 0.471 & 0.089 & 0.339 & 0.22 & 0.47 \\
TS5 & Pre & 0.28 & 0.26 & -0.079 & -0.205 & -0.205 & -0.30 & -0.30 \\
TS5 & Post & 0.60 & 0.62 & 0.492 & 0.395 & 0.395 & 0.60 & 0.60 \\
TS6 & 2024 (reference) & 0.48 & 0.65 & 0.520 & 0.353 & 0.353 & 0.45 & 0.45 \\
TS7 & 2024 (reference) & 0.48 & 0.57 & 0.403 & 0.271 & 0.271 & 0.40 & 0.40 \\
TS2 & HD-MLC (context) & 0.16 & -0.02 & -0.194 & -0.359 & -0.359 & -0.70 & -0.70 \\
TS3 & HD-MLC (context) & 0.23 & 0.01 & -0.189 & -0.419 & -0.119 & -0.72 & -0.42 \\
\bottomrule
\end{tabular}
}
\end{table}

For TS1, calibration increased the feeler-gauge gap, sweeping-gap x-offset, Stakitt gap deviation, and XML-derived MPC gap by 0.20, 0.24, 0.485, and 0.499~mm. The corresponding TS5 changes were 0.32, 0.36, 0.571, and 0.600~mm. The post-adjustment sweeping-gap values were both 0.62~mm, between TS7 (0.57~mm) and TS6 (0.65~mm), and the Stakitt values of 0.471 and 0.492~mm were likewise within the TS6/TS7 range of 0.403--0.520~mm. Restoring the configured 0.25~mm offset moved the TS1 XML-derived MPC gap to 0.339~mm, also within range.

Agreement with the reference machines was not uniform across metrics. The adjusted machines were brought within the local reference range on the sweeping-gap and Stakitt metrics, and TS1 also on the offset-restored XML-derived MPC gap, but finished slightly beyond it on the feeler gauge (0.55 and 0.60~mm against 0.48~mm on TS6 and TS7), on the offset-restored CSV-derived gap, and, for TS5, on the XML-derived MPC gap, 0.042~mm above the upper reference value. TS5 was the wider of the two on every metric except the sweeping gap, on which they finished equal. This follows from the endpoint chosen during the calibration session, a compromise among the measurements then available rather than a match to any single machine.

\subsection{Response to the applied gap change}
\label{sec:applied_response}

Because the magnitude of the service-mode adjustment was known, each method's response could be compared with a predicted value rather than assessed only for direction. The displayed gap setting was increased by 0.05~cm on TS1 and 0.06~cm on TS5, corresponding to opposing-leaf opening changes of 0.50 and 0.60~mm at isocentre. The vendor specifies the setting at the isocentre plane, so the applied values are used without projection, and the same convention is assumed for the centreline offset. The feeler gauge measures at the leaf plane and was expected to change by 0.510 times the isocentric value, while the RayStation x-offset, being per leaf end, was expected to change by half the opposing-gap change. Only the Stakitt and MPC predictions follow from the reported quantity alone, so agreement for the other two is evidence for the measurement combined with its assumed scaling. The predicted responses were therefore 0.255, 0.250, 0.500, and 0.500~mm on TS1 and 0.306, 0.300, 0.600, and 0.600~mm on TS5 for the feeler-gauge, sweeping-gap, Stakitt, and MPC metrics respectively.

All four methods used to guide and confirm the calibration recovered the applied change to within 0.060~mm. The XML-derived MPC metric agreed most closely, with residuals of 0.0013 and 0.0003~mm on TS1 and TS5, respectively. The CSV-derived MPC gap was the one metric that did not recover the applied change, overshooting by 0.450~mm on TS1 and 0.300~mm on TS5, factors of 1.9 and 1.5, despite coming from the same acquisitions. Section~\ref{sec:mpc_csv_vs_xml} discusses a possible cause. For the centreline adjustment, TS1 received no correction and its MPC-derived metric changed by only 0.008~mm, whereas the +0.02~cm correction applied to TS5 produced a change of 0.023~cm in magnitude.

\subsection{Centreline offset, reproducibility, and gap uniformity}
\label{sec:centreline_results}

The TS5 MLC centring field showed an asymmetric horizontal feature before adjustment, indicating leaf positions not centred on the collimator rotation axis. After applying the +0.02~cm centerline correction that asymmetry was substantially reduced, though not eliminated (Supplementary Figure~\suppref{fig:ts5_centreline}{S1}). This visual interpretation was consistent with the quantitative Stakitt bank-offset result, and MPC history suggested the offset had been present since installation. No centreline correction was applied to TS1, TS6, or TS7. The conventional picket-fence test repeated on TS5 after adjustment reported no change in its relative-position metric, despite the applied 0.60~mm opposing-leaf opening change detected by the other methods.

Before adjustment the two-direction TS5 offsets differed between Bank~A (+0.295~mm) and Bank~B (\(-0.090\)~mm), giving a derived centreline metric of +0.193~mm. Afterwards the offsets were similar (\(-0.234\) and \(-0.161\)~mm) and the metric was \(-0.036\)~mm. The change was consistent in direction with the reduction in asymmetry seen in the independent EPID image, although it should not be read as an absolute physical displacement, because MPC retains a machine-specific expected MLC gap value.

Leaf-position reproducibility was stable across both adjustments, the opposing-gap-equivalent value changing by less than 0.005~mm, and leaf-to-leaf gap uniformity was likewise largely unchanged, so the corrections shifted the mean positions without materially changing approach-direction dependence (Supplementary Section~S1). Neither quantity was targeted by the corrections applied here, and both warrant their own baseline and trending rather than being read as outcomes of this study.

\subsection{CSV- and XML-derived MPC gaps}
\label{sec:mpc_csv_vs_xml}

The two MPC gap metrics are not independent: they come from the same acquisitions, share the same configured target, and differ only in how each leaf's two approach-direction measurements are combined, the XML-derived metric taking their mean and the vendor export the signed value of larger absolute magnitude. Apart from export rounding, their difference is therefore attributable to that rule, and is systematic rather than random, because magnitude selection biases each leaf value away from zero and that bias reverses as a bank is driven through zero between two compared states. Both adjusted machines were calibrated across that point, and the resulting swings in the difference between the two metrics, 0.451~mm on TS1 and 0.300~mm on TS5, are almost exactly the amounts by which the CSV-derived metric overstated the applied change in Section~\ref{sec:applied_response}. The overstatement is thus not noise but a predictable consequence of tracking a change through zero with a statistic biased away from it. The derivation, the per-bank displacements, and their agreement with half the reported backlash are given in Supplementary Section~S1.

Cross-machine ordering is nevertheless largely preserved, the two offset-restored metrics correlating at \(r=0.982\) across the six Millennium-MLC observations. The exported summaries therefore remain serviceable for ranking machines or trending one machine while its bank offsets keep a constant sign, but do not support reading the size of a change from the difference between two of them.

\subsection{Exploratory cross-method associations}
\label{sec:cross_method}

\begin{figure}[H]
\centering
\includegraphics[width=\textwidth]{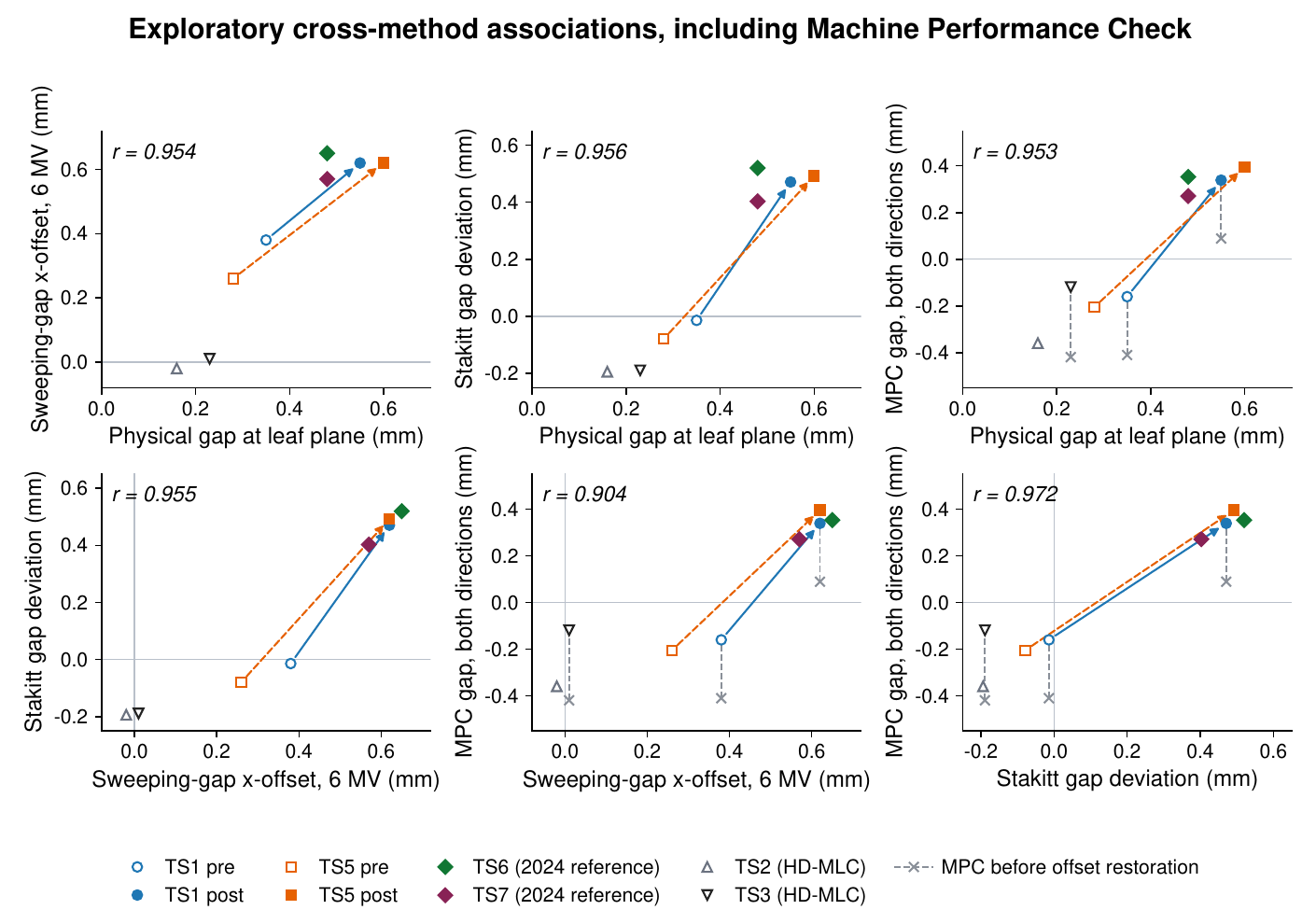}
\caption{Pairwise associations among the feeler-gauge, 6~MV sweeping-gap, Stakitt Fence, and offset-restored MPC gap metrics; the MPC gap is the XML-derived metric. Coloured arrows are the paired changes, filled diamonds are the local reference machines TS6 and TS7, separated by colour, and open triangles the HD-MLC contextual observations TS2 (upward) and TS3 (downward). Grey crosses are the as-measured MPC values, joined by dashed lines to the coloured points after the configured MPC offset is restored. Annotated Pearson coefficients span all eight observations.}
\label{fig:mlc_method_comparison}
\end{figure}

Figure~\ref{fig:mlc_method_comparison} presents the six pairwise comparisons among the feeler-gauge, 6~MV sweeping-gap, Stakitt, and offset-restored XML-derived MPC gap metrics; the corresponding correlation coefficients for both MPC derivations are given in Supplementary Section~S1.

The consistently positive associations and same-direction paired changes are compatible with the four methods responding to the underlying gap-calibration state. Across all eight observations the correlations among feeler gauge, sweeping gap, and Stakitt were 0.954, 0.956, and 0.955. Restoring the configured MPC offset increased the correlation of the XML-derived metric with every independent metric, from 0.871 to 0.953 for feeler gauge, 0.863 to 0.904 for sweeping gap, and 0.944 to 0.972 for Stakitt, with the same pattern over the six Millennium-MLC observations alone. 

\begin{figure}[H]
\centering
\includegraphics[width=\textwidth]{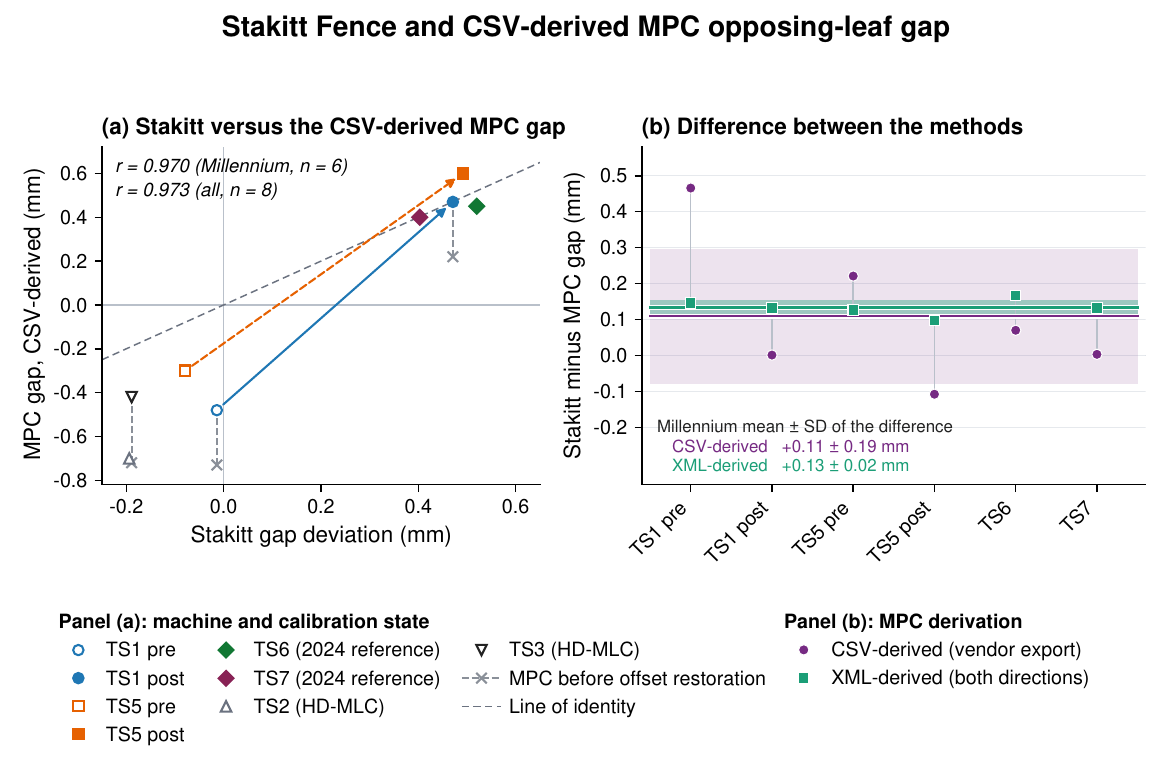}
\caption{Stakitt Fence gap against the CSV-derived MPC opposing-leaf gap. (a) Offset-restored CSV-derived MPC gap against the Stakitt value; symbols follow Figure~\ref{fig:mlc_method_comparison}, and the dashed diagonal is the line of identity, drawn for scale rather than as an expected relationship. (b) Stakitt minus MPC gap for the six Millennium-MLC observations, for both MPC derivations. Colour identifies the derivation in this panel rather than the machine, and the horizontal bands give the mean and one population standard deviation of each difference.}
\label{fig:stakitt_vs_mpc_csv}
\end{figure}

Because Barnes et al.\ compared Stakitt with MPC's standard magnitude-selected leaf-offset exports~\cite{Barnes2025CDC}, Figure~\ref{fig:stakitt_vs_mpc_csv} pairs the corresponding CSV-derived gap with the Stakitt values of Table~\ref{tab:gap_measurements} for direct comparison with that study, where Figure~\ref{fig:mlc_method_comparison} uses the XML-derived metric.

Panel (a) shows the two metrics remain strongly associated across the cohort when the CSV-derived quantity is used, at \(r=0.970\) over the six Millennium-MLC observations and \(r=0.973\) over all eight, against 0.996 and 0.972 for the XML-derived metric (Supplementary Section~S1). Panel (b) shows the association is nevertheless carried by the cross-machine spread rather than by agreement at any one observation. The Stakitt-minus-MPC difference is stable for the XML-derived metric, spanning 0.097 to 0.167~mm at a mean and population standard deviation of \(+0.134 \pm 0.021\)~mm, but ranges from \(-0.108\) to \(+0.466\)~mm for the CSV-derived metric, at \(+0.109 \pm 0.188\)~mm, and changes sign within the cohort. The two paired calibration states are the clearest instance: TS1 and TS5 each cross the point at which the magnitude-selection rule reverses, so the CSV-derived difference moves by 0.465 and 0.329~mm between pre- and post-adjustment, against 0.014 and 0.029~mm for the XML-derived difference. Both quantities plotted are opposing-leaf gaps formed from two bank values, so these differences are not directly comparable in magnitude with the per-bank differences of Barnes et al. Supplementary Section~S1 develops the candidate mechanisms for a residual offset between the two tests and the limits on reading this figure against their measurement design.

\subsection{Post-calibration model-sensitivity and clinical QA assessment}
\label{sec:arccheck_results}

\begin{figure}[H]
\centering
\includegraphics[width=0.94\textwidth]{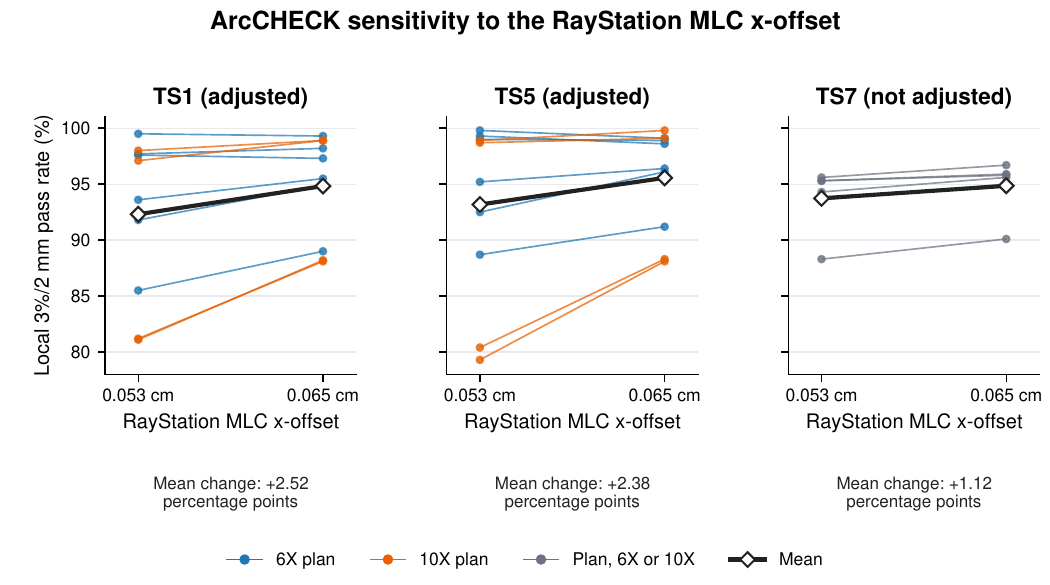}
\caption{Paired local 3\%/2~mm ArcCHECK pass rates calculated with RayStation MLC x-offsets of 0.053 and 0.065~cm. Lines join each plan's results and diamonds show machine means. TS1 and TS5 show the ten post-calibration clinical plans. TS7 shows the six separate commissioning plans of Section~\ref{sec:ts7_crosscheck}, drawn in one neutral colour because their per-plan energy assignment was not recorded. In every panel the measured dataset is fixed and only the calculation differs.}
\label{fig:xoffset_sensitivity}
\end{figure}

Mean pass rates increased for all five gamma analysis criteria when the RayStation MLC x-offset was increased from 0.053 to 0.065~cm (Table~\ref{tab:xoffset_sensitivity}). At 3\%/2~mm the mean increase was 2.52 percentage points on TS1 and 2.38 on TS5, and 15 of the 20 machine-plan pairs improved (Figure~\ref{fig:xoffset_sensitivity}). All five that did not improve were 6~MV plans already passing at 97\% or above at 0.053~cm, where the criterion is close to saturation; every pair below that level improved.

\begin{table}[H]
\centering
\caption{ArcCHECK local-gamma sensitivity to the RayStation MLC x-offset. Values are mean pass rates (\%) for the same ten clinical plans on each machine; \(\Delta\) is the paired mean change from 0.053 to 0.065~cm.}
\label{tab:xoffset_sensitivity}
\small
\begin{tabular}{lrrrrrr}
\toprule
& \multicolumn{3}{c}{TS1} & \multicolumn{3}{c}{TS5} \\
\cmidrule(lr){2-4}\cmidrule(lr){5-7}
Criterion & 0.053~cm & 0.065~cm & \(\Delta\) & 0.053~cm & 0.065~cm & \(\Delta\) \\
\midrule
3\%/3~mm & 96.89 & 97.95 & +1.06 & 97.06 & 98.14 & +1.08 \\
3\%/2~mm & 92.31 & 94.83 & +2.52 & 93.18 & 95.56 & +2.38 \\
3\%/1~mm & 71.56 & 76.51 & +4.95 & 79.61 & 84.89 & +5.28 \\
2\%/2~mm & 89.20 & 92.43 & +3.23 & 89.66 & 93.01 & +3.35 \\
2\%/1~mm & 63.33 & 69.10 & +5.77 & 72.28 & 77.53 & +5.25 \\
\bottomrule
\end{tabular}
\end{table}

\begin{figure}[H]
\centering
\includegraphics[width=0.94\textwidth]{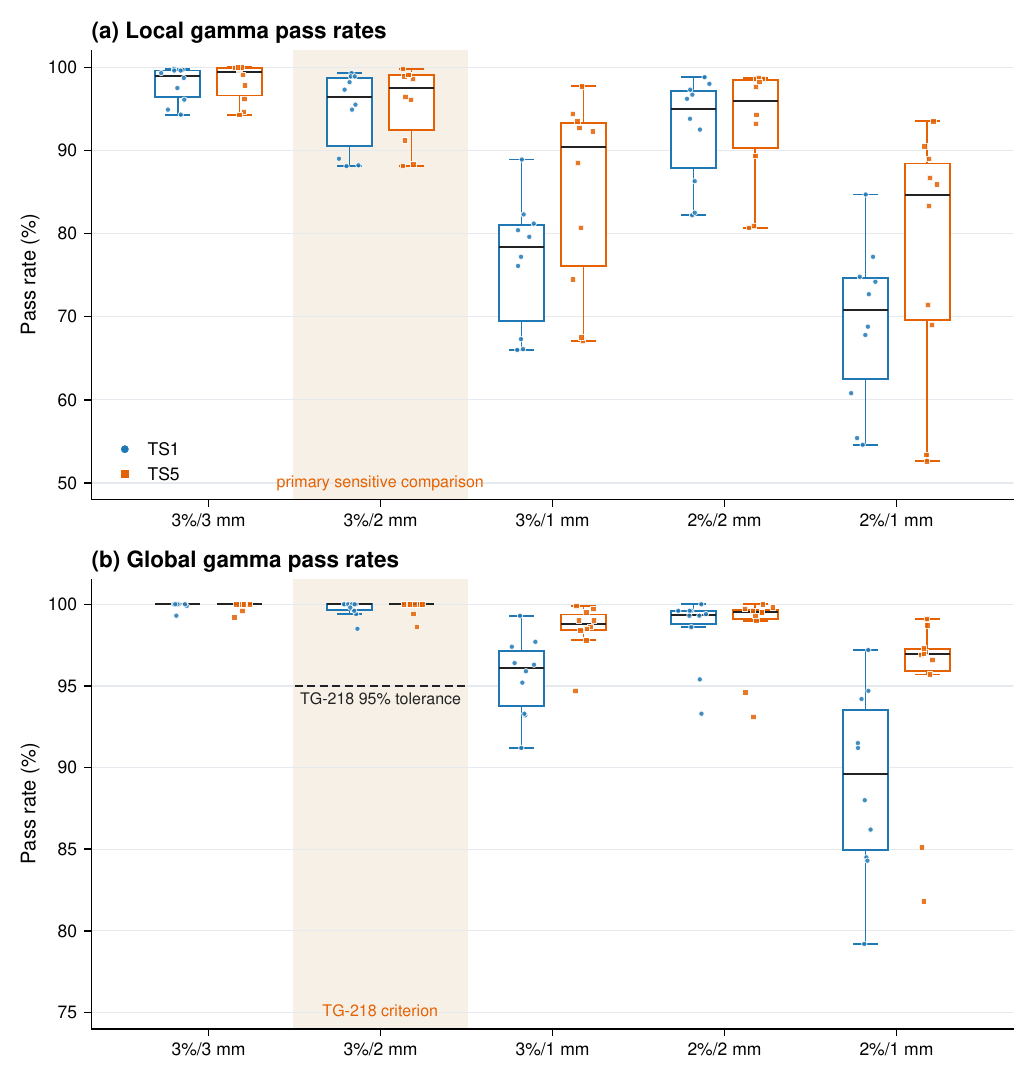}
\caption{Post-calibration (a) local-gamma and (b) global-gamma pass-rate distributions for the ten clinical plans across the five criteria. Boxes span the interquartile range, centre lines show medians, whiskers extend to the most extreme observation within 1.5 times that range, and points show all plans. The shaded column is 3\%/2~mm; the dashed line in panel (b) marks the TG-218 95\% tolerance limit.}
\label{fig:arccheck_gamma}
\end{figure}

At the sensitive local 3\%/2~mm criterion the mean pass rate for the retained model, was 94.8\% (range 88.1--99.3\%) for TS1 and 95.6\% (88.1--99.8\%) for TS5. The global 3\%/2~mm pass rates averaged 99.7\% and 99.8\%, and every clinical-plan result exceeded the TG-218 95\% tolerance limit, with minima of 98.5\% and 98.6\% respectively~\cite{Miften2018}. Figure~\ref{fig:arccheck_gamma} shows the full local and global distributions across all five criteria for each machine. Both fall as the distance-to-agreement is tightened, but the local distributions are consistently wider and lower: at 3\%/2~mm they span 88.1--99.8\% across the two machines against a global spread confined to the top 1.5 percentage points, and at the tightest 2\%/1~mm criterion the local medians fall to 70.8\% on TS1 and 84.6\% on TS5, against 89.6\% and 97.0\% globally.

\subsection{Measured cross-check of the common model on TS7}
\label{sec:ts7_crosscheck}

The six clinical plans measured on TS7 during that machine's commissioning validation were analysed against dose calculated with the pre-harmonisation model at 0.053~cm and with the final common TB\_SD model at 0.065~cm, the measured arrays being common to both, so only the calculated distribution differed. Agreement was better under the common model for every plan at every criterion, that is for all 24 plan-criterion pairs. Mean local pass rates rose from 89.8\% to 91.8\% at 2\%/2~mm and from 93.7\% to 94.9\% at 3\%/2~mm, and the corresponding global means by 0.82 and 0.65 percentage points. Every plan exceeded the TG-218 global 3\%/2~mm tolerance limit under both models, the minima being 96.1\% and 97.3\%~\cite{Miften2018}, so the acceptance conclusion is unchanged and the comparison shows a consistent increase in margin. Under the common model the mean local 3\%/2~mm pass rate on TS7 was 94.9\%, against 94.8\% on TS1 and 95.6\% on TS5. The plan sets differ between machines, so this is not a matched comparison and no difference should be read as a machine effect, but it places the unadjusted reference machine in the same range as the two adjusted machines under the model they now share.

\FloatBarrier

\section{Discussion}

\subsection{Principal findings}

The four beam-matched Millennium-MLC machines occupied a binomial calibration-state at installation. Because beam matching specifies dosimetric beam-model quantities rather than mechanical MLC state, machines intended to share a model should first have their gap calibration compared against a beam-axis-referenced or physical standard.

The study does not demonstrate that installation year caused the baseline separation. Machine age, service history, software version, prior calibration practice, and site are all potential confounders, and site is completely confounded with the observed separation. A difference in commissioning or servicing practice between the two sites would produce exactly the pattern reported here. TS5 was installed in 2021, after the reported workflow change, yet showed the narrow-gap state, so it possibly did not receive the additional step.  In any case, installation year alone does not account for the grouping, favouring the calibration actually applied as the operative variable without separating it from site or local installation practice. A published precedent points the same way: Xue et al.\ attributed unusually large measured dosimetric leaf gaps at another centre to its running TrueBeam software earlier than version~2.0, before a change they describe Varian as having made to the MLC calibration technique, which on their account produced lower values thereafter~\cite{Xue2018}. That attribution is one possible explanation rather than established, but two changes of calibration practice moving the gap-sensitive parameter in opposite directions are harder to reconcile with machine age than with the procedure a machine received at installation.

\subsection{Comparison of measurement methods}
\label{sec:method_comparison}

The four methods interrogate related but non-identical quantities. Stakitt and MPC are both derived from EPID images of MLC-defined fields, but are referenced to different expected values and implemented differently in their analysis, so a residual method-specific offset is to be expected. They nevertheless behaved as measurements of a common underlying quantity when the MPC gap was taken from the leaf-resolved arrays despite coming from different acquisitions. That distinction places the comparison of Barnes et al.\ on a common footing, since their CDC values were the magnitude-selected export quantities~\cite{Barnes2025CDC}; read on those terms the two studies are consistent, both showing that close agreement in a detected change can coexist with a non-zero method-specific correction factor. What the present data adds is that part of such a correction (or difference to expected MLC gap) can arise inside MPC rather than between the two tests.

These metrics may be compared in trend and change but not in absolute value, a caution applying equally to the feeler-gauge and sweeping-gap metrics (Table~\ref{tab:gap_measurements}). The two-direction leaf-resolved MPC metric agreed most closely with the applied changes, but this is not evidence that MPC is the most accurate: it rests on two paired observations without repeats and is in part expected by construction, since the adjustment displaces the physical leaves within the machine's own coordinate system while the MPC target is unchanged. It demonstrates internal consistency of leaf positioning as read by MPC, which is weaker than independent verification of the physical gap.

Three configuration dependencies limit MPC's use as a standalone check. First, the exported bank summaries are magnitude-selected statistics displaced away from zero by up to half the mean leaf backlash, so they exaggerated the deviations and inflated the measured response to adjustment (Section~\ref{sec:mpc_csv_vs_xml}). That rule is not stated in the vendor documentation and was established here only by reconstructing the export from the leaf-resolved arrays (Supplementary Section~S1). Barnes et al.\ report the same documentation gap~\cite{Barnes2025CDC}. Direction averaging of the leaf-resolved arrays avoids this selection bias and is preferable when estimating mean gap changes, while retaining MPC's other measurement and configuration dependencies. Second, the version 3.0 and 4.0 guides specify that the MPC gap setting contain only the accumulated dosimetric gap adjustment, excluding the installation calibration~\cite{VarianMPC30,VarianMPC40}. Because the configured value displaces the target, updating both concurrently returns the reported deviation to its previous value and can conceal a real change in opposing-leaf opening. An independently maintained local record of gap adjustments is therefore essential, and the broader need for independent checks and controlled correction factor management in manufacturer-integrated quality control applies here~\cite{Pearson2025}. Third, the default thresholds are wide, 1~mm for the mean and maximum leaf offset of each bank and 0.5~mm for leaf-position reproducibility, and all eight acquisitions passed on every exported MLC summary value (Supplementary Section~S1). MPC therefore passed TS5 in its pre-adjustment state, which carried a 0.4~mm bank misalignment and an opposing-leaf opening 0.60~mm narrower than the state to which it was subsequently calibrated, and passed TS1 with an opening 0.50~mm narrower. A pass at the default thresholds is compatible with the calibration differences this study set out to characterise, so a local action level set against the machine's own reported value is needed if MPC is to register them at all.

That the conventional picket-fence check registered no change (Section~\ref{sec:centreline_results}) is expected when leaf positions are assessed relative to the mean picket location, since a change common to both banks leaves the relative pattern intact. Routine QA intended to detect absolute calibration drift should therefore reference picket locations to known positions, or use a beam-axis-referenced method such as Stakitt. The Stakitt measurements here were made at gantry \(0^\circ\), appropriate for comparing calibration state, but a complete routine programme may require additional gantry angles given the gantry-dependent backlash effects reported by Barnes et al.~\cite{barnes_backlash}.

\subsection{Implications for treatment planning system modelling}
\label{sec:tps_modelling}

The fitted RayStation x-offset values illustrate why mechanical calibration and TPS modelling should be considered together. An effective x-offset can compensate dosimetrically for a mechanically different gap, but a shared beam model then risks representing some machines better than others. The TPS vendor sets the same expectation: having published MLC parameter sets it advises adopting as a whole, the RayPhysics manual singles out this one parameter, noting that the offset ``is still expected to vary somewhat between LINACs, even though the variation is decreasing with modern linear accelerators''~\cite{RayPhysicsManual}. That is why the local fit reported here tunes this parameter alone against the published set (Supplementary Section~S2), and it is what the mechanical harmonisation was intended to remove. Where plans transfer between machines without recalculation, mechanical harmonisation is preferable to per-machine x-offset values, provided the adjustment is supported by vendor procedures and appropriate commissioning checks. The alternative, demonstrated by Guan et al., is to retain separate models and show the residual dose difference between them is acceptable~\cite{Guan2024}.

After harmonisation the unconstrained fitted x-offsets spanned 0.057--0.065~cm at 6~MV and 0.060--0.072~cm at 10~MV, so the common TB\_SD value of 0.065~cm lay within that local envelope, though above the 0.046 and 0.050~cm Millennium values reported by Saez et al.~\cite{Saez2020} (Supplementary Section~S2). A residual remains whose sign depends on machine and energy, the largest being 0.008~cm per leaf end against TS7 at 6~MV and 0.007~cm against TS6 at 10~MV, approximately 0.16 and 0.14~mm of opposing-leaf aperture and of opposite sense. Further fine-tuning of the TS6/TS7 calibration could reduce the residual, but in our assessment is not warranted at this magnitude for the conventionally fractionated Millennium-MLC workload reported here. The same reasoning would not transfer to stereotactic treatment, for which the gap-sensitive parameter has been shown to drive dose-calculation error on an HD-MLC~\cite{Kim2018} and, on a Millennium MLC, to require optimisation against the treatment site rather than a sweeping-gap measurement alone~\cite{Middlebrook2017}.

\subsection{Recommended workflow for matching the MLC gap between linacs}
\label{sec:recommended_workflow}

We propose the following sequence, using complementary methods to establish the mechanical setting, assess dosimetric matching, and monitor constancy.

\begin{enumerate}
\item \textbf{Feeler gauge, for the initial mechanical setting.} Approximately match physical gaps at the leaf plane using the vendor service procedure. This direct, beam-free measurement is fast enough to repeat between adjustments, but is subjective and limited by blade thicknesses. The adjusted machines finished furthest from the reference range on it, so use it to establish the starting point rather than the endpoint. Consistent with vendor guidance, open the narrower gaps rather than close the wider one to avoid driving opposing leaves towards collision.

\item \textbf{Sweeping-gap measurement, to fix the dosimetric gap.} Use synchronous and asynchronous sweeping gaps~\cite{Saez2020} to determine a common gap-sensitive TPS parameter, rather than absorbing each machine's mechanical state into a separate value. A persistently outlying fit after step~1 identifies a residual difference the feeler gauge did not resolve.

\item \textbf{Stakitt Fence, to confirm calibration and monitor constancy.} Confirm opposing-leaf gap and bank offset with an absolute, beam-axis-referenced EPID measurement~\cite{Barnes2025}, then retain it as the monthly MLC constancy check. Stakitt recovered the applied changes to within 0.030~mm without assumed scaling and provides a reference for cross-machine calibration comparisons; a conventional picket-fence test is not a substitute (Section~\ref{sec:method_comparison}).

\item \textbf{MPC, for frequent constancy checks only.} Use MPC as a fast constancy check against a locally recorded correction factors (MPC gap offset), and not on its own to set or compare an absolute gap between machines, for the reasons set out in Section~\ref{sec:method_comparison}.
\end{enumerate}

\subsection{Clinical significance}

The clinical importance of a gap-calibration difference depends on aperture size, modulation, delivery technique, and how far the model compensates for the mechanical state, with small stereotactic apertures and highly modulated VMAT fields the most sensitive. Consistent with that expectation, the same measured arrays agreed better with the wider-gap 0.065~cm model on both adjusted machines and on all six independently acquired TS7 plans, and the improvement grew as the distance-to-agreement criterion was tightened. Local gamma exposed plan-dependent differences largely hidden by global normalisation (Figure~\ref{fig:arccheck_gamma}), whereas every global 3\%/2~mm result exceeded the TG-218 tolerance limit~\cite{Miften2018}. Because no matched pre-calibration baseline exists on any machine, these results do not isolate the causal effect of the adjustment. A further qualification applies to any array-based conclusion of this kind. Across the SEAFARER cohort submissions using the ArcCHECK returned sensitivities to introduced delivery errors spanning 0--100\% and specificities spanning 30--100\%, the variation tracking the analysis protocol rather than the detector~\cite{May2026}. The pass rates above therefore characterise the two models under one local protocol, and are not a general statement about the sensitivity of array measurement to gap-calibration error.

\subsection{Centreline calibration and related observations}

The study also identified a systematic centreline-offset error on TS5 that, on the available MPC history, appeared to have been present since installation. The qualitative EPID centring image, quantitative Stakitt bank-offset result, and two-direction MPC-derived centreline metric moved consistently after correction, supporting the half-difference of the bank-local offsets as a screening metric for an error the centring image alone can only show qualitatively. The local Winston--Lutz collimator-isocentre diameter also decreased. Neither this nor the MPC comparison was collected as a controlled repeated endpoint, so both are hypothesis-generating. That the offset had not been identified by routine relative-position QA reinforces the need for at least one absolute, beam-axis-referenced MLC test in the QA programme.

\subsection{Clinical implementation}
\label{sec:clinical_implementation}

After measurement review and clinical assessment, the common RayStation TB\_SD model and corresponding RadCalc models were commissioned, legacy machine-specific models deprecated, and plan-comparison and charting configurations updated to permit assignment to any of the four beam-matched machines.

The released configuration was subsequently audited externally. The Australian Clinical Dosimetry Service performed on-site Level~II audits of TS7 and TS5 on 3 and 4~June 2025, about four and a half months after the calibration work described here. Both machines were audited using TB\_SD at both energies across 3D conformal, IMRT, and VMAT deliveries, and both returned a Pass (Optimal Level) outcome, the highest of three grading categories, for every scored case and energy. This is the only evidence reported here not acquired in-house, and it covers one adjusted machine, one never-adjusted machine, and both sites. It establishes little about the MLC calibration specifically. However, the case designed expressly to verify MLC modelling is reported but not scored, the audits assessed only the retained model and only on two of the four machines, and a pass establishes that delivered dose agreed with planned dose within the service's acceptance criteria rather than that the mechanical calibration is correct. The audit design, the scored and unscored results, and these qualifications in full are given in Supplementary Section~S3.

\subsection{Limitations}
\label{sec:limitations}

This was a single-service investigation with four primary machines across two sites, only two machines in each installation-era group, and no random sampling of calibration states, so the site, installation-year, and calibration-state groupings cannot be separated from one another. Pre/post observations from TS1 and TS5 were paired and not independent, and the exploratory coefficients combine those paired states with two HD-MLC observations, so clustering may strengthen the apparent associations.

The feeler-gauge data comprised a single measurement set per machine state and the method is inherently subjective, so a rigorous uncertainty is difficult to assign. No Stakitt acquisition was repeated within a calibration state either, so no local repeatability estimate is available for the method on which the absolute cross-machine comparison rests. The published short-term repeatability on a TrueBeam, 0.02~mm at three standard deviations over five consecutive deliveries at gantry \(0^\circ\)~\cite{Barnes2025}, characterises the published implementation rather than the in-house one used here. Because the MLC harmonisation was completed within a four-day window, no ArcCHECK measurements were acquired immediately before the mechanical change, so its dosimetric effect could not be isolated. The 0.053 versus 0.065~cm analysis instead tests two calculations against the final measurement set. The TS7 cross-check likewise carries constraints: its measurements were acquired at commissioning and not repeated, so it isolates the model change exactly but cannot detect any change in TS7 itself. The six plans were assembled for commissioning validation rather than selected for modulation, and it tests the retained value against one alternative rather than a bracketing range, establishing 0.065~cm as the better of the two settings examined, not the best available.

Further methodological limitations qualify the derived quantities themselves. The RayStation fits were not accompanied by repeated acquisitions, confidence intervals, or a covariance analysis, so the constrained sensitivity fits are consistent with possible parameter coupling rather than proof of near-degeneracy, and published parameter sets cannot be treated as universal targets. 

For MPC, software version was confounded with installation year, and although the pairing of sub-acquisitions was fixed by exact reconstruction of the vendor export, which member of each pair corresponds to which physical direction of leaf travel is not identified in the vendor documentation, so the sign of any direction-specific quantity is assigned by inference and cross-machine MPC comparisons should be regarded as provisional. Service-mode adjustment also carries mechanical and clinical risks, and the procedure described here should not be generalised outside vendor guidance, local risk assessment, and qualified service and medical-physics oversight.

\subsection{Future work}

Prospective testing should combine controlled gap and centreline perturbations with repeated acquisitions, multiple feeler-gauge operators, multiple gangle measurements, and a locked MPC correction factor and configuration history. Retaining exported summaries and leaf-resolved arrays would distinguish magnitude-selection effects from between-test differences. A multi-institutional cohort is needed to separate installation-era effects from machine age and service history, and an HD-MLC pair to assess applicability to SRS/SABR.

\section{Conclusions}

The beam-matched machines had different baseline MLC calibration states: the 2024 machines showed wider effective gaps, and TS1/TS5 adjustment moved all four measurement methods consistently towards the local TS6/TS7 range. We recommend complementary use of the feeler gauge for initial mechanical setting, sweeping gaps for a common gap-sensitive TPS parameter, Stakitt for absolute gap and bank-offset verification and routine constancy, and MPC for frequent monitoring of deviation from expected MLC positions. MPC alone cannot establish absolute cross-machine gap matching. Quantitative MPC assessment should use leaf-resolved arrays rather than exported bank summaries, with an independent calibration-change record.

The post-calibration comparison and separate TS7 dataset favoured the retained RayStation x-offset. However, the primary assessment reused plans that informed model selection, and TS7 tested only one alternative. These findings concern one service, four Millennium-MLC machines and two adjustments, with calibration state confounded by site and installation era. Prospective repeated measurements and larger multi-institutional and HD-MLC cohorts are needed before generalising the workflow.

\section*{Supplementary material}
The separate Supplementary Material documents MPC extraction and derivation, a worked vendor-export example, bank-resolved CSV/XML comparisons, the Stakitt/CSV-derived MPC pairing used by Barnes et al.~\cite{Barnes2025CDC}, and cross-method correlations (S1); the fitted RayStation MLC-model parameters (S2); and the external Level~II audit of the released configuration (S3). Supplementary tables and figures are numbered S1 onwards, independently of the section numbering.

\section*{Acknowledgements}
Thank you to the medical physics and radiation engineering teams at the Royal Adelaide Hospital and the Lyell McEwin Hospital for their assistance with the MLC calibrations on the linacs. Thank you to Varian for the useful discussions about the MLC gap and offset calibration process. Thank you to Jordi Saez for the very useful discussions about his method for determining optimal RayStation MLC model parameters using synchronous and asynchronous sweeping gap measurements, and for providing the original DICOM plans.

\section*{Ethics and consent}
This work was undertaken as part of routine clinical medical physics quality assurance within the SA Health Radiation Oncology service. Retrospective treatment plans were re-delivered on linear accelerators and measured using ArcCHECK to assess MLC calibration and machine matching. No patient-specific medical imaging data were used, and no identifiable patient information is reported.

CALHN Research Services determined that the work did not constitute human research and therefore did not require review or approval by a Human Research Ethics Committee. The publication received Publication Endorsement from the CALHN Research Office Expedited Review Panel. This endorsement relates to institutional approval for publication and does not constitute ethical approval.

\section*{CRediT authorship contribution statement}
\textbf{Michael Douglass:} Conceptualization, Methodology, Software, Formal analysis, Investigation, Data curation, Visualization, Supervision, Writing -- original draft.
\textbf{Corey Bridger:} Investigation, Writing -- review \& editing.
\textbf{Mitchell Herrick:} Investigation, Writing -- review \& editing.
\textbf{Joshua Southwell:} Investigation, Writing -- review \& editing.
\textbf{Andrew Kennedy:} Investigation, Writing -- review \& editing.
\textbf{Michael Barnes:} Writing -- review \& editing.
\textbf{Joerg Lehmann:} Writing -- review \& editing.

\section*{Declaration of generative AI and AI-assisted technologies in the writing process}
During the preparation of this work the authors used Anthropic Claude and OpenAI ChatGPT to assist with writing the analysis code, and with reviewing and drafting the manuscript. After using these tools the authors reviewed and edited the content and take full responsibility for the content of the publication.

\section*{Funding}
No funding to declare.

\section*{Data and code availability}
Patient data is not available for sharing. Measurement data and code are available from the corresponding author upon reasonable request.

\section*{Conflicts of interest}
The authors have no conflicts of interest to declare.

\begingroup
\small
\bibliographystyle{elsarticle-num}
\bibliography{references}
\endgroup

\end{document}